\documentclass[fleqn,twoside]{article}
\usepackage[headings]{espcrc2}

\readRCS
$Id: espcrc2.tex,v 1.2 2004/02/24 11:22:11 spepping Exp $
\usepackage{epsfig}

\usepackage[figuresright]{rotating}

\newcommand{\AmS}{{\protect\the\textfont2
  A\kern-.1667em\lower.5ex\hbox{M}\kern-.125emS}}

\newcommand{\ord}{{\cal O}}
 
\title{Semileptonic and radiative $B$ decays circa 2005}

\author{Paolo Gambino\address[MCSD]{
 INFN, Sez.\ di Torino and Dip.\ di Fisica Teorica,
Universit\`a di Torino,   10125 Torino, Italy}
        \thanks{Supported in part by the EU grant
MERG-CT-2004-511156 and by MIUR under contract 2004021808-009.}}
 
\begin{document}

\begin{abstract}
I briefly review  the theoretical 
status of semileptonic and radiative $B$ decays in 2005.

\vspace{1pc}
\end{abstract}

\maketitle

\section{Introduction}
Semileptonic and radiative beauty decays play an important role in the 
determination of the CKM matrix elements and in the indirect search
for new physics: the former, tree-level dominated, allow for a precise
measurement of the CKM elements $V_{cb}$ and  $V_{ub}$; 
the latter, loop-induced,
are directly sensitive to new physics contributions, and give also information
on $V_{td}$ and $V_{ts}$. These decay modes, all characterized by an 
electroweak current that probes the $B$ dynamics, have a lot in common and
form a remarkable set of interdependent measurements. 
Their simplicity, however, is only apparent:
if one is interested in precision measurements they display all
the rich complexity specific of QCD dynamics.
The main theoretical divide runs between inclusive and exclusive decays:
inclusive decays can be studied using an Operator Product Expansion (OPE), 
parameterizing the non-perturbative physics in terms of 
$B$-meson matrix elements of power-suppressed local operators, 
while exclusive decays require an estimate of the form factors using 
non-perturbative methods (lattice QCD, sum rules).
Having two completely different methods at our disposal represents a huge
advantage and makes non-trivial cross-checks possible. See \cite{yb} for a 
review.

\section{Exclusive determination of $V_{cb}$}
 The exclusive determination of $|V_{cb}|$ employs the extrapolation of
the $B\to D^* l \nu$ rate to the kinematic endpoint where the $D^*$ is produced
at rest (zero-recoil). In this limit, the form factor $F(1)$ is known, up
to corrections suppressed by at least two powers of $\Lambda_{QCD}/m_{c,b}$ 
that have to be
computed, e.g.\ on the lattice. Since one needs to estimate only the
$\ord(10\%)$ correction to the heavy quark limit, a good  accuracy
can be reached even with present non-perturbative
methods. In fact, current lattice QCD and
sum rule results are both consistent with $F(1)=0.91\pm0.04$ \cite{yb}.
The overall uncertainty is therefore close to 5\%: $|V_{cb}^{excl}|=41.2(1.0)_{
ex}(1.8)_{th}\times 10^{-3}$, but the two most precise experimental results, by
Babar and Cleo, differ by almost 3$\sigma$ \cite{hfag}.
Semileptonic decays to $D$ mesons give consistent but less precise results.
Progress is expected especially from unquenched lattice calculations.

\section{Inclusive determination of $V_{cb}$}
While the non-perturbative unknowns in the exclusive determination
of $|V_{cb}|$ have to be calculated, those entering the
inclusive semileptonic decay, $B\to X_c l \nu$, can be measured in a
self-consistent way. Indeed, the inclusive decay rate depends only on
the hadronic structure of the decaying $B$ meson, but the sensitivity to it is
actually suppressed by  two powers of $\Lambda_{QCD}/m_b$, as the
highly energetic decay products are (generally) unable to probe the long
wavelengths characteristic of the $B$ meson. Formally, an 
OPE allows us to write the differential $B\to X_c l \nu$ rate 
as a double series in $\alpha_s$ and $\Lambda_{QCD}/m_b$,
whose leading term is  nothing but the parton model result. However, the OPE
results for the spectra can be compared  to experiment only after smearing
over a range of energies $\gg\Lambda_{QCD}$ and away from the endpoints.
The hadronic mass spectrum, for instance, is
dominated by resonance peaks  that have no counterpart in the OPE:
the OPE results have no {\it local} meaning.

The observables that can be studied in the OPE include the total rate
and moments (weighted integrals) of the lepton energy and hadronic mass
spectra, as well as the photon spectrum in radiative decays.
They generally are subject to a lower cut on the
charged lepton energy and can be written in a way 
analogous to that of the integrated rate,
\begin{eqnarray}
\Gamma_{cl\nu}&=& \frac{G_F^2 m_b^5 \eta_{ew}}{192\pi^3} |V_{cb}|^2  z(r)\left[
1+a_1(r) \frac{\mu_\pi^2}{m_b^2} \right.\nonumber\\
&&
\left.\! \! \! \! 
+a_2(r) \frac{\mu_G^2}{m_b^2}+
b_1(r) \frac{\rho_D^3}{m_b^3}+b_2(r) \frac{\rho_{LS}^3}{m_b^3}
+...\right],\nonumber
\end{eqnarray}
where $r=(m_c/m_b)^2$, $z(r)$ is the tree-level expression,
$\eta_{ew}$ contains the leading electroweak 
corrections,
the Wilson coefficients $a_i,b_i$ are series in $\alpha_s$,
and power corrections up to $1/m_b^3$ have been kept. 
The parameters entering the predictions 
are  $\alpha_s$,  properly defined {\it quark }
masses $m_{c,b}$, and  the $B$ meson matrix elements of the four {\it local}
operators that appear up to $O(1/m_b^3)$: 
$\mu_{\pi,G}^2$ and $ \rho^3_{D,LS}$.
Because they depend on the various parameters in different ways,
the moments serve a double purpose: they allow to
constrain the non-perturbative parameters and they test the overall
consistency of the OPE framework. 
Effects that cannot be described by the OPE
(and therefore violate {\it parton-hadron duality})  and higher order power
corrections can be severely constrained.
                                                                               
Recent experimental results \cite{momexp}, analyzed
in the light of up-to-date theoretical predictions \cite{Gambino,Bauer,Oliver},
have led to a big step forward, both in completeness and  accuracy. 
There is 
a remarkable consistency of a variety of leptonic and hadronic moments,
leading to an excellent fit. The values of the quark masses are
in agreement with
lattice and spectral sum rule determinations, and the
other non-perturbative parameters are determined for the first time at the
10-20\% level, in agreement with theoretical expectations.
 Radiative moments from Belle, Cleo, and Babar \cite{radmom}
can be included as well
without deteriorating the quality of the fit, that yields
$|V_{cb}^{incl}|=41.58(0.45)_{ex}(0.58)_{th}\times 10^{-3}$ \cite{Oliver}, 
in agreement with the exclusive result.
The estimate of the theory error (missing higher order perturbative and 
non-perturbative contributions, intrinsic charm etc.) 
is  particularly delicate; different recipes in different schemes
\cite{Gambino,Bauer}
have led to compatible results. Despite recent progress \cite{pert},
higher order perturbative corrections to the Wilson coefficients are  
the main source of uncertainty: a 1\% determination of $|V_{cb}|$ is possible
but requires new calculations.
It should be stressed that the OPE parameters describe {\it universal}
properties of the $B$ meson and of the quarks. For example, 
$m_b$ and  $m_b-m_c$ are determined in the fit 
within less than 40 and 30 MeV, resp. The reach of the
new results therefore  extends  well beyond the $|V_{cb}|$ determination, 
as is well demonstrated by the case of $|V_{ub}|$.

\section{Exclusive determination of $V_{ub}$}
The ratio $|V_{ub}/V_{cb}|$ measures the left side of the unitarity triangle,
identifying a circle in the $(\bar\rho,\bar\eta)$  plane.
The determination of $|V_{ub}|$
from $b\to u $ semileptonic decays parallels that of $|V_{cb}|$, but the
exclusive determination ($B\to \pi l\nu, B\to \rho l\nu$, etc.)
is penalized by the absence of a heavy quark
normalization for the form factors at a certain kinematic point.
 Moreover, if  theoretical precision is lower, so is
statistics, by about two orders of magnitude. 
\begin{figure}[t]
\vspace{-.4cm}
\epsfig{file=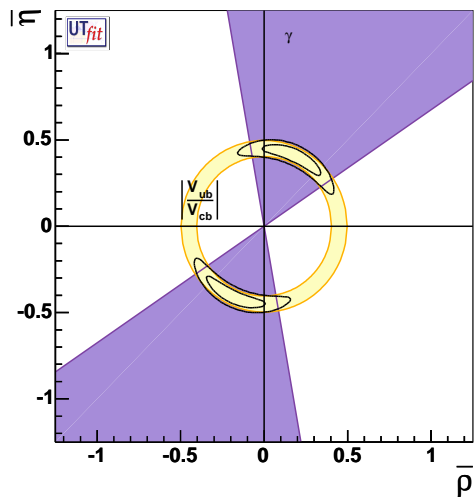,width=7.5cm}
 \caption{\it
      Determination of the Unitarity Triangle using only tree level processes,
$|V_{ub}|$ and $\gamma$ from $B\to D K$.
    \label{exfig} }
\end{figure}
In view of the precision reached by $|V_{cb}|$, a drastic improvement in the
determination of $|V_{ub}|$, made possible by the high statistics available,
has become  the top priority. The relevance of $|V_{ub}|$ is 
is illustrated for instance in Fig.~1, where the Unitarity Triangle 
is determined for the first time using tree-level
processes only. Comparing a high precision determination of this kind, 
insensitive to new physics, to the standard one based on loop processes 
would be very instructive \cite{uut}.
 
In the exclusive case, lattice QCD and light cone sum
rules complement each other, but as the first unquenched calculations
appear, the error in the high-$q^2$ region accessible to lattice
still  exceeds 10\%, while the $q^2$ extrapolation is well under control,
thanks to analyticity and new experimental data \cite{becher}. 
Sum rules prefer a lower value than lattice,  $|V_{ub}|= (3.2\pm 0.1\pm 0.1
\pm 0.3)\times 10^{-3}$ (first of \cite{becher}) against  
 $|V_{ub}|=(4.1\pm 0.6)\times 10^{-3}$ \cite{hfag,latticevub}. 
The goal of lattice simulation is a 5-6\% determination within a few years 
\cite{ckmvub05}. Recent proposals include a new $q^2=0$ 
form factor normalization based on SCET  \cite{arnesen}, and  
the combination of $B\to K^* l^+l^-$ and 
$B\to \rho l\nu$ data \cite{grinstein}. The latter is 
not yet competitive and the $|V_{ub}|$ result 
could depend on new physics in the rare decay.

\section{Inclusive determination of $V_{ub}$}
The inclusive determination of $V_{ub}$ is strongly affected by the
kinematic cuts necessary to isolate  $b\to u $ transitions from the dominant
$b\to c$ background. In general,  cuts placed near the perturbative
singularities (typically, the lepton energy endpoint)
destroy the convergence of the OPE and introduce a sensitivity to
local $b$-quark wave function properties like the Fermi motion of 
the heavy quark in the $B$ meson. These non-perturbative
effects are  not suppressed by powers of $1/m_b$: at leading order in 
this expansion they are described by a single 
distribution function, often called {\it shape function}, whose first
moments
are given by expectation values of the same {\it local} operators we have 
encountered earlier. The shape function 
 is universal:  in principle it can be extracted from the photon spectrum 
in $B\to X_s \gamma$ or
studying the differential distributions in $B\to X_u l \nu$, although the two 
processes are different at subleading order in $1/m_b$ and $\alpha_s$. 
The shape function gets renormalized by perturbative effects: disentangling
the latter from non-perturbative contributions is not trivial and has been
done in different ways \cite{sfren,Benson:2004sg}.

Different cutting strategies have been proposed: 
cuts on the hadronic invariant mass $M_X<M_D$, on the electron energy, 
on the $q^2$ of the lepton pair and
combinations thereof, on the light-cone variable $P_+=E_X-|\vec P_X|$; 
each  has peculiar experimental and theoretical
systematics \cite{luke}, though the uncertainty on leading and subleading 
shape functions plays often a central role. 
Eventually, the variety of complementary approaches that have been developed
will be extremely useful.
Recent  theoretical work  has focused on 
the optimization of the cuts, subleading non-perturbative effects \cite{subsf},
the resummation of Sudakov logs, the role of the  radiative decay
spectrum in  constraining  the shape function, etc. \cite{recent}.

The latest HFAG average of inclusive determinations,
$|V_{ub}|=4.38(33)\ 10^{-3}$ \cite{hfag} agrees with the exclusive 
determination based on lattice. The error  is close to 8\%  and 
dominated by theoretical systematics. 
The progress since last year is significant and due to: 
i) the implementation of the constraints on $m_b$ and on the moments of the
shape function derived from $b\to c l\nu$ spectral moments;
ii) larger statistics and better knowledge of the charm background,
  that has allowed to cut in a milder way, to the relief  of theorists.

Further improvement can be expected from high statistics data. 
A more precise determination of the radiative spectrum
and better experimental constraints on the Weak Annihilation (WA) 
contributions \cite{WA} are needed.
Particularly promising are new analyses based on 
fully reconstructed events that allow high discrimination of 
charmed final states. They allow the  measurement of 
$b\to u$ decay distributions well beyond the kinematic cuts on 
$b\to c$ \cite{bu_moments}. Of course for  milder cuts the 
experimental error tends to increase, while the sensitivity to the 
shape function decreases: the balance between theoretical and 
experimental errors can be optimized. 
The $b\to u$ differential distributions and their truncated
moments will help constraining the shape function(s), 
Weak Annihilation (WA) contributions, and the heavy quark parameters
\cite{ossola}.

\section{Radiative decays}
While the V-A current involved in the semileptonic $B$ decays is conserved,
that is not the case of the tensor current that induces radiative decays.
As a consequence, these decays depend logarithmically on the electroweak scale
at which the current is generated by $W$ and top quark loops:
in addition to the $b$ mass and the scale of QCD, 
a third scale $\sim M_W\gg m_b \gg \Lambda_{QCD}$ must be taken 
into account. The large logs $L= \log M_W/m_b$ are  resummed 
in the context of an effective theory, at leading order in  $1/M_W$: 
the resummation of $O(\alpha_s^n L^n)$
term corresponds to the leading order (LO) expression, 
of $O(\alpha_s^n L^{n-1})$ terms to the NLO, etc.
Apart from this complication, and from additional ones due to the presence of
charm loops \cite{charm}, the OPE for {\it inclusive} radiative 
decays is analogous to the one for semileptonic ones. 
The NLO calculation was practically completed in 1996 
\cite{chetyrkin} and involves
$O(\alpha_s)$ matrix elements. Electroweak effects and power corrections 
are also known \cite{ew,charm}.
The main theoretical uncertainty in the NLO 
analysis of the Branching Ratio (BR) 
is related to the mass of the charm quarks 
circulating in the $O(\alpha_s)$ loops of the  matrix element 
$\langle X_s \gamma | (s_L c_L)(c_L b_L) | b \rangle$ \cite{gambino-misiak}. 
This matrix element vanishes at LO, and the charm dependence 
is an $O(\alpha_s)$ effect. The {\it natural} scale
at which $m_c$ should be normalized is therefore undetermined without 
and $O(\alpha_s^2)$ calculation. 
%Even though 
%the pole mass $M_c\sim 1.5$~GeV appears inappropriate in this case,  
The matter is quite relevant:
the numerical difference between $m_c(m_c)\simeq 1.25$~GeV and 
$m_c(m_b)\simeq 0.85$~GeV is large enough to shift the BR by more than 10\%.
Using $m_c(m_b/2)$ as central value, the BR of $B\to X_s \gamma$ for 
$E_\gamma>1.6$~GeV is  $(3.60\pm 0.30)\times 10^{-4}$, with
an error close to 8\% \cite{gambino-misiak}, dominated by the charm scale 
uncertainty. In practice, a lower cut on the photon energy 
($E_\gamma<1.8$-1.9~GeV) is always applied to avoid background. 
A detailed knowledge of the tail of the spectrum is therefore required for the 
extrapolation to a region where  the OPE can be trusted.
The same problem also affects the moments of the photon spectrum that are 
employed in the HQE fit \cite{Benson:2004sg}.
Both perturbative hard gluon emission and the tail of the shape function 
concur to form the tail of the photon spectrum. As already mentioned, 
disentangling them is not straightforward. 
The transition from the shape function dominated region to the local OPE
has been studied by Neubert, who found that it
can be described by a multi-scale non-local OPE  \cite{neubert}. 
He noticed that the cut $E_\gamma<E_{cut}$ effectively introduces 
two new scales that may be
relevant besides $m_b$: $\Delta= m_b-2E_{cut}$ and $\sqrt{m_b \Delta}$. 
According to this picture, the perturbative tail of the spectrum 
receives contributions at the scale $\Delta$, and may be subject to large  
higher order perturbative corrections, even for $E_{cut}\le 1.8$~GeV, simply 
because $\Delta$ is then close to 1~GeV. 
This view has been criticized  \cite{kolya}. Certainly, 
the results of a new calculation of the dominant $O(\alpha_s^2)$ 
effects in the photon spectrum \cite{melnikov}
show only very small  deviations from the 
LO plus BLM perturbative spectrum considered in \cite{Benson:2004sg}, 
excluding large corrections in the standard picture.
Moreover, the partial BR calculated in the multiscale OPE,
BR($E_\gamma>1.6$~GeV)= $(3.47^{+0.33}_{-0.40}$$^{+0.32}_{-0.29}
)\times 10^{-4}$ \cite{neubert} has a 
central value very close to that of \cite{gambino-misiak}. 
The values of  $m_b$ and $\mu_\pi^2$ 
extracted from the first and second photonic moments 
in \cite{Oliver} following \cite{Benson:2004sg} are also very close 
to those extracted from the same data  
using the multiscale OPE \cite{neubertnew}. 
One would conclude that the local OPE is likely
to provide a good description of the spectrum in the intermediate region
$E_\gamma\sim 1.6-1.8$~GeV of phenomenological interest, and that there is
no need to enlarge substantially the theoretical error on the partial BR
given in \cite{gambino-misiak}.

\begin{table*}[htb]
\caption{Summary of the main theoretical limitations.}
\label{table:1}

\newcommand{\m}{\hphantom{$-$}}
\newcommand{\cc}[1]{\multicolumn{1}{c}{#1}}
\renewcommand{\tabcolsep}{1pc} % enlarge column spacing
\renewcommand{\arraystretch}{1.2} % enlarge line spacing
\begin{tabular}{@{}|l|l|l|l|l|}
\hline
process & quantity & Th error & needs & goal \\
\hline
$B\to D^* l \nu$ & $|V_{cb}|$ & $\sim 4$\% &unquenching, analytic work
  & 1\% \\
$B\to X_c l \nu$ & $|V_{cb}|$ & $\sim 1.5$\% & new pert calculations & $<$1\% \\
$B\to \pi(\rho) l \nu$ & $|V_{ub}|$ & 10-15\% & 2-loop lattice matching etc.   & 6\% \\
$B\to  X_u l \nu$ & $|V_{ub}|$ & $\sim 6-7\%$ & more data/synergy with th & $<5\%$ \\
$B\to  X_s \gamma$ & BR & $\sim$10\% & NNLO  & $<$5\% \\
$B\to  \rho^0\gamma/B\to K^*\gamma$ 
& $|V_{td}/V_{ts}|$ & 10-20\% & lattice SU(3) breaking etc & ? \\
\hline
\end{tabular}
\end{table*}

In order to compare different experiments, the cut rates are usually
extrapolated to a conventionally defined 
{\it total} rate \cite{kagan}. The NLO result
 $(3.73\pm 0.30)\times 10^{-4}$ \cite{gambino-misiak} can then
be compared with the experimental world 
average: $3.39^{+0.30}_{-0.27}\times 10^{-4}$ \cite{hfag}. No deviation 
from the SM is observed, but in view of 
the final accuracy expected at the $B$ factories the theoretical
prediction has to be improved.  
That is the aim of the NNLO calculation currently
 under way \cite{melnikov,nnlo}. To date, 
the missing  pieces of this challenging enterprise include the four loop
anomalous dimension matrix and the finite parts of the three loop 
matrix elements with charm loop.

The high precision of inclusive radiative $b\to s$ decays 
is not yet matched by the theoretical understanding of 
{\it exclusive} radiative $B$ decays \cite{yb}, 
despite some recent progress \cite{rad_excl}. Because of the  
difficulty of measuring inclusively $b\to d\gamma$, 
the ratio of $b\to d\gamma$ over $b\to s\gamma$ exclusive modes 
is extremely interesting. $B\to (\rho,\omega) \gamma$ has just been 
measured for the first time \cite{bdg_exp}.
The ratio $B\to \rho^0 \gamma/B\to K^*\gamma$, in particular, 
is hardly affected by WA contributions, while the SU(3) breaking effects
can be estimated on the lattice and using light cone sum rules; the
error is 10-20\% at most \cite{boschrad}. In this way, we can 
extract $|V_{td}/V_{ts}|$ before $\Delta M_s/\Delta M_d$ is measured.
While the 2004 preliminary result showed an interesting deviation from the 
global UT fit, the recent BELLE update has restored consistency with the SM
and damped enthusiasm  \cite{bdg_exp,utfit}.

Finally,  it should  also be mentioned that the  
rare leptonic transitions $b\to s  l^+ l^-$  complement 
    radiative decays in constraining new physics \cite{bsll1}. 
The inclusive decay 
$B\to X_s l^+ l^-$, in particular, has reached experimental and 
    theoretical maturity with theoretical errors comparable to 
$B\to X_s\gamma$ \cite{bsll2}.

\section{Conclusions}
The joint theoretical and experimental 
effort to study  semileptonic and radiative decays 
to high precision has led to relevant progress in the 
determination of the CKM matrix elements and in testing the Standard Model.
 Table 1 summarizes the present theoretical uncertainty 
and outlines the main ingredients necessary to improve on it for the various 
processes. While no deviation from the Standard Model has been so far 
uncovered, more theoretical work is needed 
to make the most of the wealth of data coming from the $B$ factories.

\vskip .4cm
  
I am grateful to M.~Misiak for a careful reading of the manuscript, to
O.~Buchm\"uller and H.~Fl\"acher for useful communications, and to
the organizers of Beauty 2005 for the invitation.

\end{document}